\documentclass[aps,prb,twocolumn,color,pdflatex]{revtex4}

\usepackage{graphicx}% Include figure files
\usepackage{dcolumn}% Align table columns on decimal point
\usepackage{amsmath,amssymb,bbold,bm,epsfig}

\begin{document}

\newcommand{\bk}{{\bf k}}
\newcommand{\bp}{{\bf p}}
\newcommand{\bv}{{\bf v}}
\newcommand{\bq}{{\bf q}}
\newcommand{\bs}{{\bf s}}
\newcommand{\bmm}{{\bf m}}
\newcommand{\tbq}{\tilde{\bf q}}
\newcommand{\tq}{\tilde{q}}
\newcommand{\bQ}{{\bf Q}}
\newcommand{\br}{{\bf r}}
\newcommand{\bR}{{\bf R}}
\newcommand{\bB}{{\bf B}}
\newcommand{\bE}{{\bf E}}
\newcommand{\bA}{{\bf A}}
\newcommand{\bK}{{\bf K}}
\newcommand{\vd}{{v_\Delta}}
\newcommand{\tr}{{\rm Tr}}
\newcommand{\bj}{{\bf j}}
\newcommand{\bn}{{\bf \hat{n}}}
\newcommand{\cH}{{\cal H}}
\newcommand{\cT}{{\cal T}}
\newcommand{\cM}{{\cal M}}
\newcommand{\cG}{{\cal G}}
\newcommand{\BGamma}{{\bm \Gamma}}
\newcommand{\bsig}{{\bm \sigma}}
\newcommand{\bpi}{{\bm \pi}}

\title{Lattice model for the surface states of a topological insulator
with applications to magnetic and exciton instabilities}

\author{D.J.J. Marchand and M. Franz}
\affiliation{Department of Physics and Astronomy,
University of British Columbia, Vancouver, BC, Canada V6T 1Z1}

\begin{abstract}
A surface of a strong topological insulator (STI) is characterized by an odd
number of linearly dispersing gapless electronic surface states. It is
well known that such a surface cannot be described by an effective
two-dimensional lattice model (without breaking the time-reversal
symmetry), which often hampers theoretical efforts to quantitatively
understand some of the properties of such surfaces, including the
effect of strong disorder, interactions and various symmetry-breaking
instabilities. Here we formulate a lattice model that can be used to
describe a {\em pair} of STI surfaces and has an odd number of Dirac
fermion states with wavefunctions localized on each surface.   The
Hamiltonian consists of two planar tight-binding models with
spin-orbit coupling, representing the two surfaces, weakly coupled by
terms that remove the extra Dirac points from the low-energy
spectrum. We illustrate the utility of this model by studying the
magnetic and exciton instabilities of the STI surface state driven by
short-range repulsive interactions and show that this leads to results
that are consistent with calculations based on the continuum model as
well as three-dimensional lattice models. We expect the model
introduced in this work  to be
widely applicable to studies of surface phenomena in STIs.

\end{abstract}
\maketitle

\section{Introduction}

The physical feature that makes topological insulators special is
their surface states.\cite{mooreN,hasan_rev,qi_rev} The surface states
are topologically protected, 
have been well documented experimentally,\cite{cava1,cava2,chen1} and are predicted to exhibit a wide
range of interesting phenomena when subjected to various
perturbations.\cite{qi1,EssinMagneto,QiMonopole,ZangMonopole,rosenberg2,tse1,maciejko1,ran1,rosenberg1}
  Also, they harbor a potential for future practical
applications in spintronics,\cite{nagaosa1,garate2} low-dissipation electronics\cite{seradjeh1,nagaosa2} and
topological quantum computation.\cite{fu2,cook}

The most
interesting in this regard are the three dimensional strong topological
insulators whose surface hosts an odd number of gapless fermionic
states, while their bulk remains fully gapped. Of these, materials in
the Bi$_2$Se$_3$ family represent a canonical example with a single
Dirac state located at the center of the surface Brillouin zone. At
low energies, this state can be described by a continuum Dirac
Hamiltonian,
\begin{equation}\label{dir}
h_\bk=v(k_x s_y-k_y s_x),
\end{equation}
where $\bs=(s_x,s_y,s_z)$ are Pauli matrices in the spin space and $v$
is the characteristic Fermi velocity. The spectrum is
$\epsilon_\bk=\pm v|\bk|$ and can be gapped only by adding a term
proportional to $s_z$ to the Hamiltonian, thus breaking the
time-reversal symmetry $(\cal{T})$.

Now suppose we wish to describe this surface state by a 2D lattice
model such  that the above Dirac spectrum
would emerge at low energies. Discretizing Hamiltonian (\ref{dir}) e.g.\
on a simple square lattice we obtain a candidate Hamiltonian
\begin{equation}\label{dir_latt}
h_\bk^{\rm latt}=v(s_y\sin{k_x} -s_x\sin{k_y}),
\end{equation}
with the spectrum $\epsilon_\bk^{\rm latt}=\pm
v\sqrt{\sin^2{k_x}+\sin^2{k_y}}$.  We observe that this spectrum
indeed has a Dirac point at $\bk=(0,0)$ but additional 3 Dirac
points at $(0,\pi)$, $(\pi,0)$ and $(\pi,\pi)$ have been introduced by
the discretization procedure. While it is possible to remove these
`extra' Dirac points, e.g.\ by adding a term
$s_z(2-\cos{k_x}-\cos{k_y})$ to the Hamiltonian (\ref{dir_latt}),
the price one pays is a Hamiltonian with broken $\cal{T}$ which
therefore cannot faithfully describe the physics of the
$\cal{T}$-invariant state on the STI surface.

The above construction illustrates the constraints imposed by the
well-known Nielsen-Ninomyia theorem\cite{NN}, which states that it is
impossible, as a matter of principle, to construct a $\cal{T}$-invariant
lattice Hamiltonian with an odd number of Dirac fermions in the
low-energy spectrum. The theorem assures us that the surface states of
a STI do not have a faithful description in terms of a two-dimensional
lattice model. To describe such surface states one either must use the
continuum model Eq.\ (\ref{dir}), or, if a lattice description is
required, one must consider a 3D lattice model with open boundary
conditions along at least one spatial direction.

The above conclusion imposes a number of restrictions on the
theoretical studies of the phenomena associated with the STI surface
states. While many {\em
  qualitative} properties follow directly from the continuum effective theory
the latter often fails when quantitatively accurate predictions are
required. This is because quantities computed from the Hamiltonian
(\ref{dir})  typically depend strongly on the ultraviolet cutoff that
must be imposed in order to regularize divergences. For instance, as explained in
more detail below, the critical coupling for interaction-driven
magnetic and excitonic instability in the Hamiltonian (\ref{dir}) depends
{\em linearly} on the ultraviolet cutoff $\Lambda$, which leads to
considerable uncertainty since the latter is not well known. Also,
predictions that are strongly cutoff dependent must be viewed as
inherently unreliable. To obtain a quantitatively reliable prediction
one must resort to a 3D lattice model in a geometry with surfaces
(e.g.\ a slab). In this case, calculations are typically to be performed
numerically because of the low symmetry of the problem. Such
calculations are often difficult, essentially because the vast majority
of the computational effort is spent on the bulk states which do not show
any interesting physics but nevertheless must be included in order to
capture the correct surface physics.        

In this paper we show that it is possible to construct a simple 2D
lattice model that faithfully describes the physics of a {\em pair} of STI
surfaces, such as those terminating a slab. The idea is based on a
simple observation that a pair of STI surfaces will have an even
number of Dirac points and therefore the Nielsen-Ninomyia
theorem\cite{NN} does not prevent us from constructing a relevant 2D
lattice model. Our basic model-building strategy is straightforward: we
describe each surface by a 2D Hamiltonian very similar to Eq.\
(\ref{dir_latt}) and then construct a $\cal{T}$-invariant  coupling between the two
surfaces that preserves the gapless Dirac points at $\bk=(0,0)$ but
hybridizes, and therefore gaps out, the remaining Dirac points. This
way, we end up with a 2D lattice Hamiltonian describing a pair of
parallel STI surfaces each containing a single gapless Dirac
point. The important feature of the model is that the low-energy
eigenstates are localized on a single layer with a negligible overlap
between layers,
and therefore faithfully capture the physics of the STI surfaces.

To ascertain the validity of the proposed model we conduct a number of
tests. We show that applying various perturbations (such as inducing a
magnetic gap at the Dirac point) to a single layer only affects the
low-energy physics of that layer and leaves the other layer unchanged,
exactly as one would expect in a physical STI sample. We also calculate the
spectral function of the surface state from a full 3D model and demonstrate
that, with a judicious choice of parameters, the 2D model introduced
here gives the same result in the low-energy sector.

 Finally, as a potentially useful
application of our new model we study the mean-field phase
diagram of a thin-film STI in the presence of a short-ranged repulsive
interaction (both intra- and inter-layer) and under external
bias. This configuration exhibits two leading instabilities: a magnetic
gap associated with spontaneous $\cal{T}$ breaking and an excitonic gap,
driven by condensation of electron-hole pairs residing in
different surfaces.  Both the magnetic and the excitonic phase exhibit
interesting physical properties. The magnetic phase has been predicted
to show the half-integer quantum Hall effect,\cite{fu1} image magnetic monopole
effect,\cite{QiMonopole} and inverse spin-galvanic effect,\cite{garate2}  to name just a few, while
the exciton condensate (EC) phase is expected to show dissipationless transport of
electron-hole pairs, interesting thermo-electric properties, as well
as fractionally charged vortices.\cite{seradjeh1,peotta1,moon1,seradjeh2} In addition, the interplay between
the magnetic and EC orders has been argued to exhibit a number of
interesting phenomena, such as the anyon exchange statistics for
certain topological defects.\cite{cho1} Similarly, interfacing such EC with a superconductor is predicted to result in interesting Majorana edge modes.\cite{seradjeh3}  Using the proposed model we map out the corresponding phase diagram and study its evolution in response to the changes in parameters that can be tuned in a laboratory setting. The understanding that we thus gain will aid experimental searches for these interesting phases of quantum matter.

%%%%%%%%%%%%%%%%%%%%%%%%%%%%%%%%%%%%%%%%%%%%%%%%%%%%%%
\section{Lattice model for the surface}

The basic strategy for our model design has been already outlined in Sec.\ I. Here we provide the necessary technical details. To summarize our main objective we wish to construct a 2D lattice model whose low-energy properties will coincide with those of a surface of a realistic STI as modeled, e.g.\ by the standard 3D lattice model. To this end it is useful to visualize a STI in a slab geometry as composed of individual layers (e.g.\ the quintuple layers in Bi$_2$Se$_3$). We can now imagine integrating out the electronic degrees of freedom associated with the bulk layers, which are gapped, and retaining only the degrees of freedom residing in the outermost surface layers. Such a procedure will leave us with a lattice model for the two surface layers, coupled through the bulk degrees of freedom that have been integrated out.
\begin{figure*}[t]
\includegraphics[width = 14.0cm]{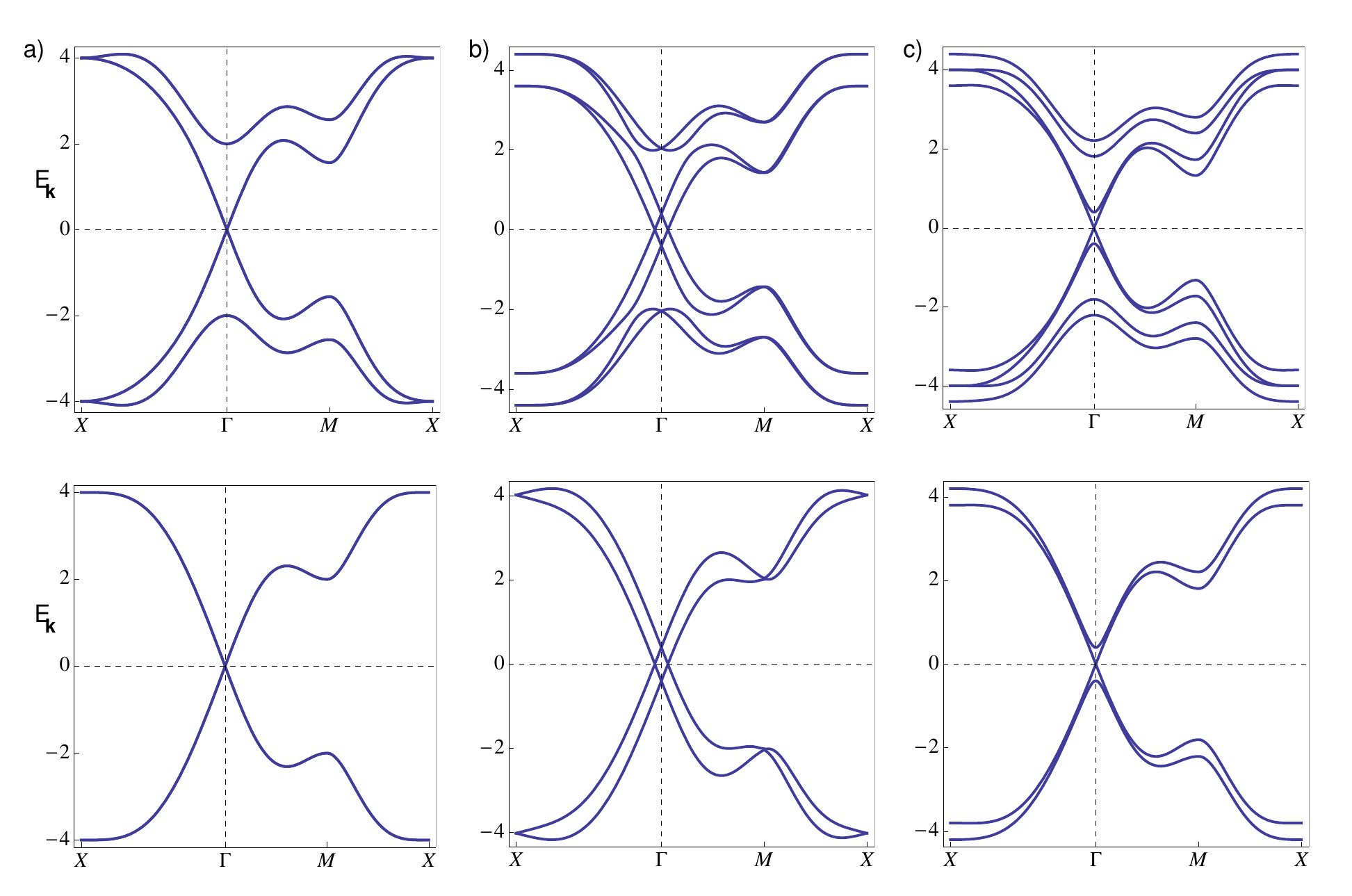}
\caption{Band structure of the  effective surface Hamiltonian for model I (top row) and model II (bottom row). Panel (a) shows the spectrum of $H_I$ (top) and $H_{II}$ (bottom) in the first Brillouin zone along the path connecting points of high symmetry $(\pi,\pi)\to (0,0) \to (0,\pi) \to (\pi,\pi)$. Panel (b) displays the effect of biasing the layers through the inclusion of $\delta H_V$ with $V=0.2$ and (c) shows the effect non-zero exchange coupling $m=0.2$ in one of the surface layers. The parameters used are $t=0.5$ and $\epsilon=2.0$. 
}\label{fig1}
\end{figure*}

Implementing this procedure in a lattice model turns out to be quite cumbersome and we defer this to a later section. Here, we proceed to construct the 2D model using a simple heuristic procedure based on the symmetries and general considerations.         

%%%
\subsection{Model I}

Our basic building block will be the effective Hamiltonian\cite{qi_rev,fu-berg1} for a single quintuple layer of Bi$_2$Se$_3$ regularized on a simple square lattice,
\begin{equation}\label{dir_latt1}
H_0=\begin{pmatrix}
h_\bk & M_\bk \\
M_\bk & -h_\bk
\end{pmatrix},
\end{equation}
where 
\begin{equation}\label{hk1}
h_\bk=2\lambda(s_y\sin{k_x} -s_x\sin{k_y}),
\end{equation}
and 
\begin{equation}\label{mk1a}
M_\bk=\epsilon-2t(\cos{k_x}+\cos{k_y}).
\end{equation}
The $2\times 2$ matrix structure in $H_0$ reflects the two orbital degrees of freedom per quintuple layer that are necessary to capture the physics of the topological phase in the Bi$_2$Se$_3$ family of materials. In the following we take $\lambda=1$ and measure all other energy scales in units of $\lambda$.
In the bulk the model parameters $\epsilon$ and $t$ have certain fixed values that are material specific. At the surface of a STI, however, we expect these parameters to be effectively renormalized so that gapless states can emerge. Considering the spectrum of $H_0$ it quickly becomes clear that the condition for the gapless states to occur at the $\Gamma$ point [i.e.\ at $\bk=(0,0)$] is $\epsilon=4t$. For future use we thus introduce 
\begin{equation}\label{mk1b}
\bar{M}_\bk=2t(2-\cos{k_x}-\cos{k_y})
\end{equation}
to denote this fine-tuned interlayer coupling.
With this choice of parameters $\bar{M}_{\bk=\Gamma}$ vanishes and $H_0$ will have two copies of a gapless Dirac fermion at this point, one associated with each orbital. If $H_0$ is to describe a single STI surface then one of these must be removed. As discussed previously, the only way to accomplish this in a $\cal{T}$-invariant manner is to introduce coupling to the degrees of freedom associated with the other surface. This consideration inspires the following $8\times 8$ Hamiltonian for the pair of surfaces,   
\begin{equation}\label{dir_latt2}
H_I=\begin{pmatrix}
h_\bk & \bar{M}_\bk & 0 & 0 \\
\bar{M}_\bk & -h_\bk & 2\Omega_\bk & 0 \\
0 & 2\Omega_\bk^\dagger & h_\bk & \bar{M}_\bk \\
0 & 0 & \bar{M}_\bk & -h_\bk
\end{pmatrix},
\end{equation}
where $\Omega_\bk$ represents the aforementioned coupling. The Hamiltonian $H_I$,  which we henceforth call `model I', acts on an 8-component wavefunction $(\Psi_1,\Psi_2)$ where $\Psi_{\ell}$ represent the orbital and spin degrees of freedom associated with surface $\ell=1,2$. There is a considerable freedom in selecting the form of the coupling between the surfaces, the only significant constraint being that $\Omega_\bk\neq 0$ near the $\Gamma$ point, so that the redundant Dirac point is indeed removed from the spectrum. Thus, $\Omega_\bk=$ const is one possible choice, although we find that taking
\begin{equation}\label{omega}
\Omega_\bk={1\over 4}[\epsilon+2t(\cos{k_x}+\cos{k_y})]
\end{equation}
leads to better results when the overall spectrum is compared to the exact surface spectrum in a 3D STI model discussed below.

The spectrum of Hamiltonian (\ref{dir_latt2}) consists of four doubly degenerate branches given by
\begin{equation}\label{ek1}
E_\bk=\pm\left[\epsilon_\bk^2+\left(\sqrt{\bar{M}_\bk^2+\Omega_\bk^2}\pm\Omega_\bk\right)^2\right]^{1/2},
\end{equation}
with $\epsilon_\bk^2=4(\sin^2{k_x}+\sin^2{k_y})$.
It is easy to see that near the $\Gamma$ point, where $\bar{M}_\bk$ vanishes, the spectrum consists of four gapless branches, comprising two Dirac points, and four gapped branches. The full spectrum is displayed in Fig.\ \ref{fig1}a.
It is also easy to see that the wavefunctions associated with the two Dirac Fermions are localized on different surfaces: near the $\Gamma$ point, where $\bar{M}_\bk$ vanishes, the two outermost diagonal elements in the Hamiltonian (\ref{dir_latt2}) become decoupled from the rest of the system and thus the corresponding eigenstates can be chosen to have support in a single surface.

It must be noted that slightly away from the $\Gamma$ point $\bar{M}_\bk$ becomes non-zero and this leads to a weak mixing of the two surface states. The strength of this mixing can be ascertained from Eq.\ (\ref{ek1}) by estimating the correction $\delta E_\bk$ to the pure Dirac dispersion $E_\bk=\pm\epsilon_\bk$ coming from the second term in the angular brackets in the limit of small $|\bk|$. Expanding the square roots for $\bar{M}_\bk^2\ll\epsilon_\bk^2\ll\Omega_\bk^2$ this correction reads
\begin{equation}\label{ek2}
\delta E_\bk\simeq {\bar{M}_\bk^4\over 8\epsilon_\bk\Omega_\bk^2}\sim k^7,
\end{equation}
and is, therefore, negligibly small for the momenta close to the $\Gamma$ point. The weak mixing of the two surface states  is an artifact of the 2D model in the sense that in a real STI slab this effect will be suppressed exponentially in the slab thickness $d$.  This discrepancy can be in principle removed by considering a different form of $\bar{M}_\bk$ that would more strongly vanish in the neighborhood of the $\Gamma$ point. Such a form is easy to write in the momentum space but would necessarily imply longer range inter-surface hopping in the real space. Since our objective is to propose and test a simple model we shall not pursue this issue here, although we point out some consequences of this weak mixing when warranted.   

A simple but instructive test of the model can be conducted by adding to Hamiltonian (\ref{ek1}) a term 
\begin{equation}\label{HdV}
\delta H_V={\rm diag}(V,V,-V,-V).
\end{equation}
Physically this corresponds to introducing an opposite bias for the two surfaces which can be achieved e.g.\ by placing the sample inside a capacitor. Such a bias will play an important role in our considerations  of the exciton condensate later in the paper. The spectrum of the modified Hamiltonian is displayed in Fig.\ \ref{fig1}b. We observe that while the high-energy bands remain unchanged the Dirac points at low energies are simply moved up and down in energy, exactly as one would expect for the degrees of freedom localized in a single layer.

Another interesting modification consists of a replacement $h_\bk\to h_\bk+m s_z$. This represents a magnetic perturbation with a moment along the $z$ direction (i.e.\ perpendicular to the surfaces). Fig.\ \ref{fig1}c shows the spectrum that results from applying such a perturbation to one of the two surfaces. We observe that, as expected, one of the Dirac cones acquires a gap while the other remains gapless.

%%%
\subsection{Model II}
In some situations it will be desirable to simplify the surface Hamiltonian one step further. Specifically, if we are only interested in the low-energy surface degrees of freedom we may want to integrate out the four gapped bands appearing in the spectrum of $H_I$. A quick reflection suggests that a very simple Hamiltonian of the form 
\begin{equation}\label{dir_latt77}
H_{II}=\begin{pmatrix}
h_\bk & \bar{M}_\bk \\
\bar{M}_\bk & -h_\bk
\end{pmatrix}
\end{equation}
will have the correct qualitative properties, if we take the two diagonal blocks to describe the two surfaces. The spectrum of this $\cT$-invariant Hamiltonian indeed has a single Dirac point at $\Gamma$ associated with each surface. The low-energy properties are very similar to $H_I$ and are illustrated in the bottom row of Fig.\ \ref{fig1}.

These considerations suggest that the heuristically constructed lattice Hamiltonians (\ref{dir_latt2}) and (\ref{dir_latt77}) indeed can be used to describe the physics of a pair of surfaces in a slab of a STI. As mentioned previously, such a description offers a range of advantages in modeling such surfaces, especially when quantitatively accurate predictions are desired. In the following Section we conduct further tests of these model Hamiltonians by comparing their spectral functions to the one derived from the exact surface propagator for the full 3D lattice model.

%%%%%%%%%%%%%%%%%%%%%%%%%%%%%%%%%%%%%%
\section{Exact surface propagator}

We now return to the idea of explicitly deriving the surface theory of a STI by integrating out the bulk degrees of freedom in a 3D lattice model. We start by writing down the action $S_0$ for the fermionic degrees of freedom in a STI slab. At finite inverse temperature $\beta=1/k_BT$ this action reads
\begin{eqnarray}\label{s1}
S_0=\int_0^\beta d\tau\{\bar{\Psi}(\partial_\tau+H_s)\Psi &+& \bar{\chi}(\partial_\tau+H_b)\chi  \nonumber\\
&+&  \bar{\Psi}T^\dagger\chi + \bar{\chi}T\Psi\},
\end{eqnarray}
where $\tau$ is the imaginary time, $\Psi$ and $\chi$ are Grassman fields representing the surface and bulk fermionic degrees of freedom, respectively.  $H_s$ is the Hamiltonian of the two surface layers, $H_b$ describes all the bulk layers and $T$ represents the terms that couple surfaces to the bulk. In the tight-binding description that we pursue in this work $H_s$, $H_b$ and $T$ should be thought of simply as matrices in spin, orbital and lattice spaces; e.g.\ $H_s$ will be given by Eq.\ (\ref{dir_latt1}) for each surface. 
We are interested in the effective surface action defined as 
\begin{equation}\label{s2}
e^{-S_{\rm eff}[\bar{\Psi},\Psi]}=\int{\cal D}[\bar{\chi},\chi]e^{-S_0[\bar{\Psi},\Psi;\bar{\chi},\chi]}.
\end{equation}
Since we are neglecting interactions at this stage the integral is elementary and we obtain
\begin{equation}\label{s3}
S_{\rm eff}=\int_0^\beta d\tau\left\{\bar{\Psi}[(\partial_\tau+H_s) - T^\dagger(\partial_\tau+H_b)^{-1}T]\Psi\right\}.
\end{equation}
For time-independent Hamiltonians it is useful to pass to the Matsubara representation by defining $\Psi(\tau)=\beta^{-1}\sum_ne^{-i\omega_n\tau}\Psi(i\omega_n)$ with $\omega_n=(2n+1)\pi/\beta$ and $n$ integer. The effective action then takes the form
\begin{equation}\label{s4}
S_{\rm eff}={1\over\beta} \sum_n\bar{\Psi}(i\omega_n)\left[\cG_s^{-1}(i\omega_n)- T^\dagger\cG_b (i\omega_n)T\right]\Psi(i\omega_n),
\end{equation}
where $\cG_s(i\omega_n)=-(i\omega_n-H_s)^{-1}$ is the bare surface propagator and $\cG_b(i\omega_n)=-(i\omega_n-H_b)^{-1}$ is the bare bulk propagator. Eq.\
(\ref{s4}) informs us that the surface degrees of freedom are described by the effective propagator $\cG_{\rm eff}$ of the form
\begin{equation}\label{g1}
\cG_{\rm eff}(i\omega_n)=\left[\cG_s^{-1}(i\omega_n)- T^\dagger\cG_b(i\omega_n)T\right]^{-1}.
\end{equation}

For a given 3D lattice model the effective surface theory can thus be obtained by performing the requisite matrix multiplications and inversions indicated in Eq.\ (\ref{g1}). While this is certainly possible to accomplish in principle, in reality one encounters at least two problems when implementing this procedure. First, for a system with many bulk layers $H_b$ and $T$ will be large matrices (for the simplest model working in momentum space for each layer $H_b$ is a $4N\times 4N$ matrix where $N$ is the number of bulk layers). Such matrices can only be inverted using numerical techniques. Second,  $\cG_{\rm eff}(i\omega_n)$ when analytically continued to real frequencies, will not have a simple form $-(\omega-H_{\rm eff})^{-1}$ and therefore, except at low energies, will not yield a simple description in terms of a hermitian Hamiltonian. It is easy to see that for $\omega>\Delta$, the bulk bandgap, $\cG_{\rm eff}$ will have a complex-valued self energy, expressing the physical fact that the surface electrons can decay into the bulk states which have been integrated out. 

%%%
\subsection{Slab geometry}

In the following we shall perform a numerical evaluation of $\cG_{\rm eff}$ in a system with a finite number of bulk layers $N$ and we also derive an analytic solution for  $\cG_{\rm eff}$ in the limit $N\to\infty$ and for special choice of model parameters. We use these results to further validate our proposed effective Hamiltonian for the TI surfaces. In these calculations we use the effective model for TIs in the Bi$_2$Se$_3$ family of materials\cite{qi_rev,fu-berg1} regularized on the simple cubic lattice. For a slab that is infinite in the $(x,y)$ plane and consists of $N+2$ quintuple layers  ($N$ bulk layers and 2 surface layers) the relevant Hamiltonian can be written as a $4(N+2)\times 4(N+2)$ matrix,
\begin{equation}\label{dir_latt4}
H=\begin{pmatrix}
H_0 & R & 0 & 0 & \dots & 0 & 0 \\
R^\dagger & H_0 & R & 0 & \dots & 0 & 0\\
0 & R^\dagger & H_0 & R & \dots & 0 & 0\\
0 & 0 & R^\dagger & H_0  & \dots & 0 & 0\\
 \vdots &  \vdots &  \vdots  &  \vdots & \ddots & \vdots & \vdots \\
0 & 0 & 0 & 0 & \hdots  & H_0 & R \\
0 & 0 & 0 & 0 & \hdots & R^\dagger & H_0
\end{pmatrix}.
\end{equation}
Here $H_0$, defined in Eq.\ \ref{dir_latt1}, describes each layer and 
\begin{equation}\label{R1}
R=-\begin{pmatrix}
0 & t_z-\lambda_z \\
t_z+\lambda_z & 0
\end{pmatrix},
\end{equation}
is the interlayer coupling, with $t_z$ and $\lambda_z$ representing the two types of interlayer hopping amplitudes permitted by the time-reversal and crystal symmetries. We note that $R$ is proportional to a unit matrix in the spin space and is therefore a $4\times 4$ matrix in the combined orbital and spin spaces. Here, and hereafter we treat the spin degrees of freedom implicitly.

The surface and bulk Hamiltonians are now easily identifiable from Eq.\ (\ref{dir_latt4}). We obtain $H_s={\rm diag}(H_0,H_0)$, $T=(R,0,0,\dots,0)^T$ and $H_b$ contains the remaining $4N\times 4N$ matrix. 

\begin{figure}[t]
\includegraphics[width = 8cm]{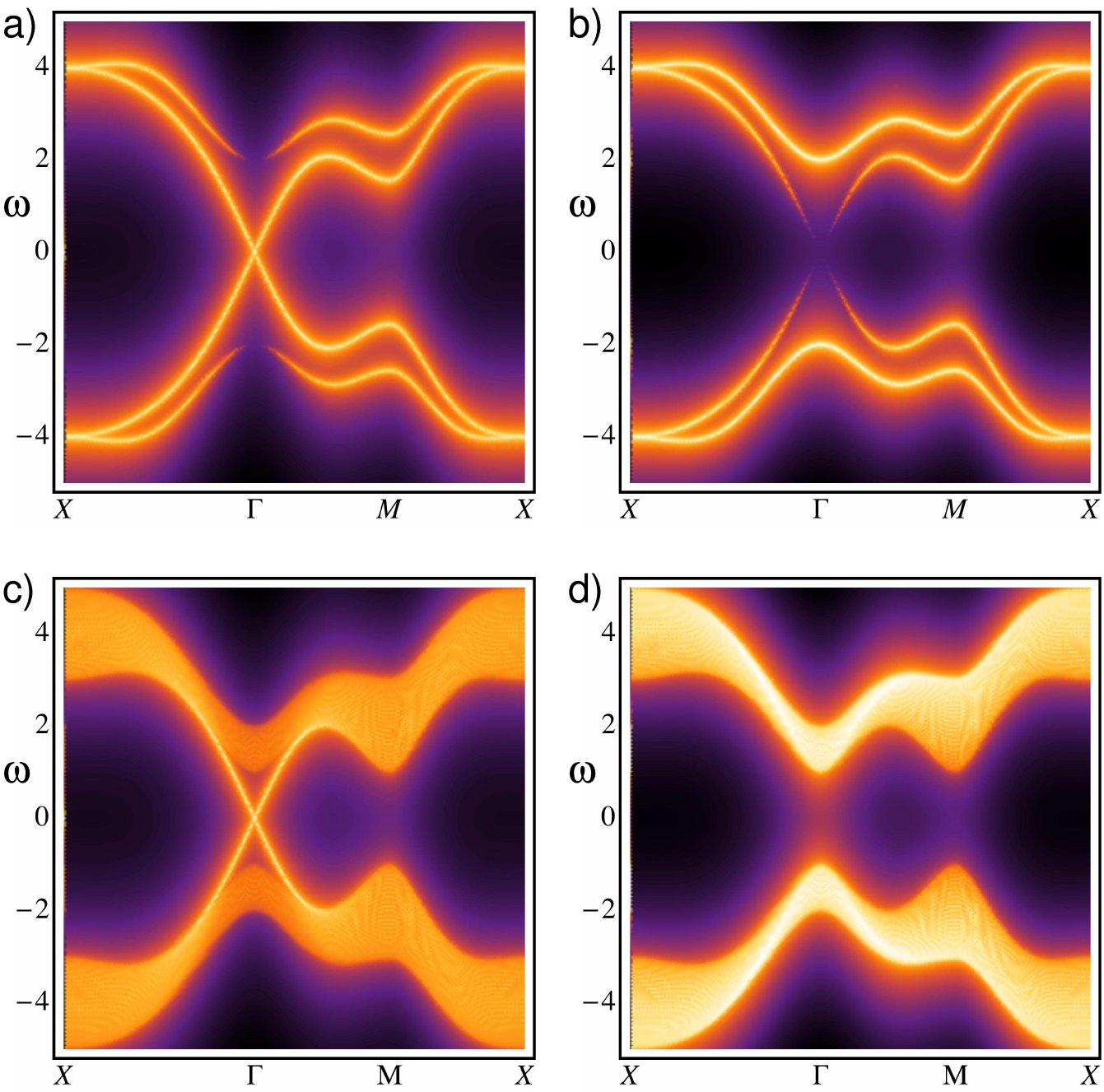}
\caption{Spectral functions $A(\bk,\omega)$ for the surface states of a model TI as calculated from model I for a) orbital 1 and b) orbital 2. Panels c) and d) show the exact spectral function for the surface layer calculated from Eq.\ (\ref{g1}) for a system with $N=46$ bulk layers and again for orbital 1 and 2 respectively. For each orbital both spin projections are included. For the bottom row we use $t=t_z=0.5$, $\lambda_z=1.0$ and $\epsilon=2.0$ corresponding to bulk STI phase with $Z_2$ index (1;000). Spectral functions are represented as density plots with light colors showing regions of high spectral density. A broadening parameter $\delta=0.02$ was used when producing these plots. 
}\label{fig2}
\end{figure}
With these ingredients it is now straightforward to numerically evaluate $\cG_{\rm eff}$ from Eq.\ (\ref{g1}) when the number of layers $N$ is not too large. Fig.\ \ref{fig2} shows the relevant spectral function $A_{\rm eff}(\bk,\omega)=-\pi^{-1}{\rm Im}\cG_{\rm eff}(\bk,\omega+i\delta)$ evaluated for $N=46$ and a set of realistic model parameters indicated in the caption.
The spectral function shows the characteristic gapless surface mode with the Dirac cone centered at the $\Gamma$ point of the Brillouin zone and a continuum of states above the bulk bandgap. As mentioned above this continuum reflects the physical fact that an electron injected into the surface at the  energy above the gap has a short lifetime due to the hybridization with the bulk states. For comparison we also show the spectral function $A(\bk,\omega)$ associated with one of the surfaces as calculated from our simple two-layer model I given in Eq.\ (\ref{dir_latt2}). The latter shows the same low-energy surface state as the exact spectral function but of course cannot reproduce the continuum of levels that exists above the gap. Nevertheless we observe that the gapped bands that follow from Eq.\ (\ref{dir_latt2}) actually mimic the spectral form of the bulk states as well as can be expected from such a simple model. We also note that the low-energy excitations in both model I and the exact spectral function derive from a single orbital. We have also verified that the electron states exhibit the correct spin-momentum locking  
expected of the surface states in a strong topological insulator.

%%%
\subsection{Semi-infinite slab}
 
As an interesting application of the formalism introduced above we now show how a simple analytic formula can be derived for $\cG_{\rm eff}$ in the limit of semi-infinite slab $N\to\infty $. This exact expression will be employed in our subsequent calculations of magnetic and exciton instabilities and can be useful potentially in other calculations.
In this case we are concerned with a single surface and thus we take $H_s=H_0$. 
To obtain $\cG_{\rm eff}$ in this situation we note that removing a single layer from the surface of a semi-infinite slab will necessarily leave the surface propagator unchanged. Since the electron hoping is only allowed between the neighboring layers it is permissible to describe the layer just below the surface by the effective propagator $\cG_{\rm eff}$.   This leads to the conclusion that $\cG_b$ in Eq.\ (\ref{g1}) can be replaced by $\cG_{\rm eff}$. Thus, for semi-infinite slab we have a self-consistent equation
\begin{equation}\label{g2}
\cG_{\rm eff}(i\omega_n)=\left[\cG_0^{-1}(i\omega_n)- T^\dagger\cG_{\rm eff}(i\omega_n)T\right]^{-1},
\end{equation}
where $\cG_0(i\omega_n)=-(i\omega_n-H_0)^{-1}$. 

In its general form Eq.\ (\ref{g2}) involves solving for a $4\times 4$ matrix $\cG_{\rm eff}$ for each value of the momentum $\bk$ and frequency $\omega_n$. One can simplify this problem by first noting that of all terms in Eq.\ (\ref{g2}) only $h_\bk$ has non-trivial dependence on electron spin ($M_\bk$ and $T$ are proportional to the identity matrix in the spin space). For each value of $\bk$ one can thus perform a rotation in spin space $h_\bk\to \tilde{h}_\bk= U^\dagger_\bk h_\bk U_\bk$ with $U=\exp{(-i{\pi\over 4}\hat{m}\cdot{\bf s}})$ and $\hat{m}$ the unit vector in the direction of $(\sin{k_x},\sin{k_y},0)$. The rotated Hamiltonian is diagonal in the spin space, 
\begin{equation}\label{hk3}
\tilde{h}_\bk=s_z\epsilon_\bk\equiv s_z2\sqrt{\sin^2{k_x}+\sin^2{k_y}}.
\end{equation}
 In this `helical' basis the problem for up and down spins decouples and we obtain two independent copies of Eq.\ (\ref{g2}) for the two spin projections at each $\bk$. The resulting equation for the $2\times 2$ matrix $\cG_{\rm eff}$ can now be solved analytically (it is, basically, a quadratic equation for a  $2\times 2$ matrix). Unfortunately, for a generic set of parameters the explicit solution is quite complicated and does not lend itself to a convenient analysis.

\begin{figure}[t]
\includegraphics[width = 8cm]{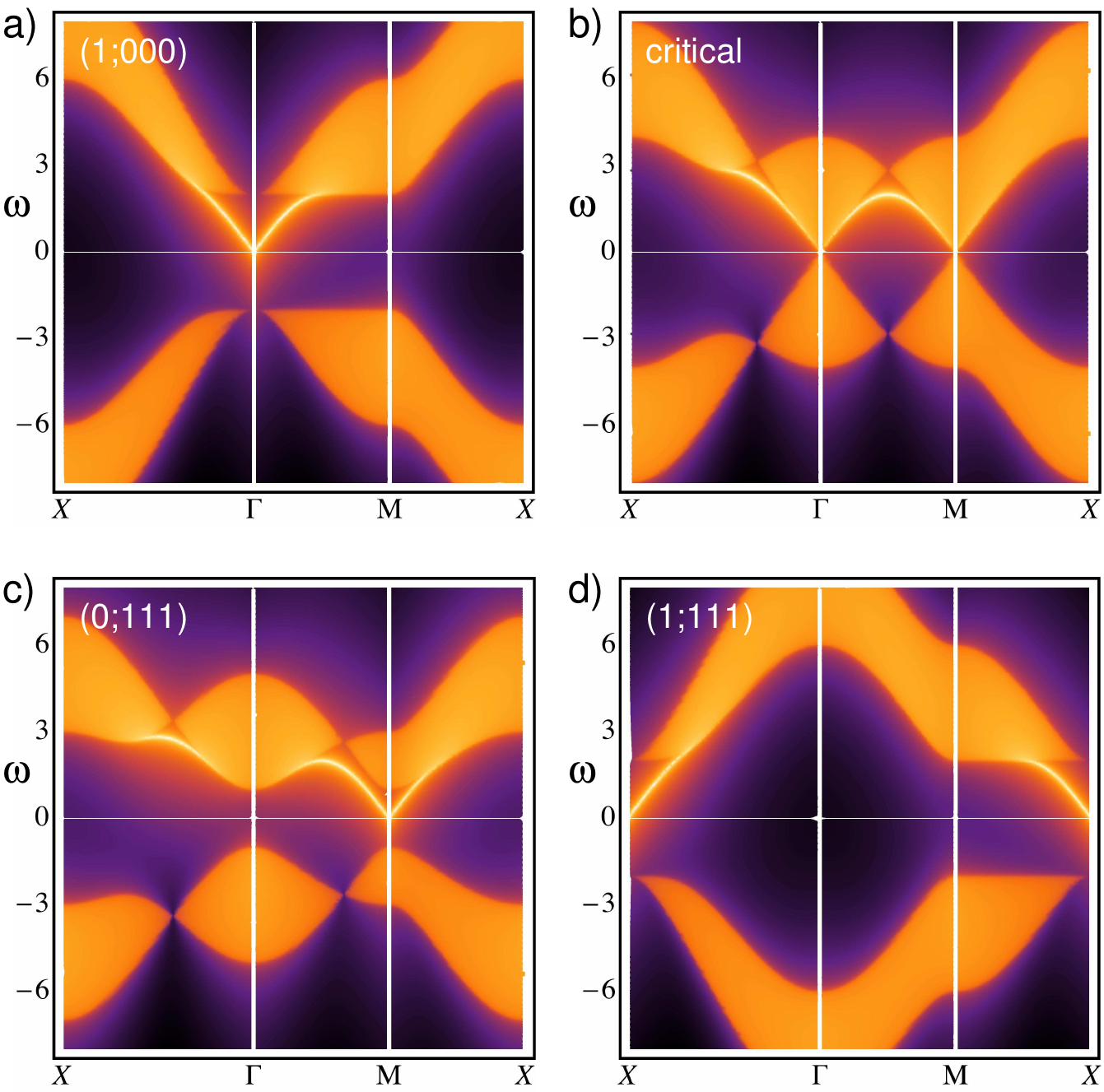}
\caption{Exact spectral functions $A^{(1)}_{\rm eff}(\bk,\omega)$ for the surface states of a semi-infinite TI slab as calculated from Eq.\ (\ref{gret}). In all panels $t_z=t=\lambda_z=1.0$ and $\delta=0.02$. Panel a) shows STI in (1;000) phase with $\epsilon=4.0$, panel b) shows a critical point at $\epsilon=2.0$, panel c) displays a weak TI in (0;111) phase for $\epsilon=1.0$ and panel d) shows a (1;111) STI with $\epsilon=-4.0$.
Note that in all plots spectral functions for a single spin helicity are shown. Those of the opposite helicity can be obtained by replacing $\omega\to -\omega$ in the existing plots.
}\label{fig3}
\end{figure}
For this reason we focus in the following on a special case defined by the condition $t_z=\lambda_z$ (but all remaining parameters arbitrary). As one can see by inspecting Eq.\ (\ref{R1}) at this special point a further decoupling in the orbital space occurs. Namely, if we define an effective surface propagator  $\cG_{\rm eff}^{(o)}$ for each of the two orbitals $o=1,2$ in the surface layer then it is easy to see that the matrix Eq.\ (\ref{g2}) breaks down to two coupled equations for scalar propagators $\cG_{\rm eff}^{(1,2)}$:
\begin{eqnarray}\label{cg12}
\cG_{\rm eff}^{(1)} &=& \left[H_+-M\cG_{\rm eff}^{(2)} M\right]^{-1}, \\
\cG_{\rm eff}^{(2)} &=& \left[H_--Q \cG_{\rm eff}^{(1)} Q\right]^{-1}. 
\end{eqnarray} 
Here $H_\pm=(-i\omega_n\pm \epsilon_\bk)$, $Q=-2\lambda_z$ and $M$ is defined in Eq.\ (\ref{mk1a}). These equations now have a simple solution,
\begin{eqnarray}\label{geff1}
\cG_{\rm eff}^{(1)}&=&{1\over 2Q^2H_+}\biggl[(H_+H_-+Q^2-M^2)\\
&&\pm\sqrt{(H_+H_-+Q^2-M^2)^2-4Q^2H_+H_-}\biggr],\nonumber
\end{eqnarray}
and similar for $\cG_{\rm eff}^{(2)}$. We shall see below that only the solution with the minus sign has the correct analytical structure and thus represents the physical surface propagator. Also, it is important to remember that $\cG_{\rm eff}^{(1)}$ represents an effective surface propagator for orbital 1 with all other degrees of freedom (including orbital 2) integrated out, and conversely for $\cG_{\rm eff}^{(2)}$.

We now study the spectral function $A_{\rm eff}^{(1)}(\bk,\omega)$ associated with the propagator (\ref{geff1}). It is particularly interesting to see how an odd number of surface states emerges from this simple propagator. Upon analytical continuation $i\omega_n\to \omega+i\delta$ we see that the denominator has poles when $\omega=\epsilon_\bk$. At low energies the only contribution to the spectral function comes from these poles (because the square root in the numerator is always real). Thus to study the surface states we can evaluate the propagator at the poles, obtaining
\begin{equation}\label{geff2}
\cG_{\rm eff}^{(1)}(\bk,\omega+i\delta)\bigg|_{\omega\to\epsilon_\bk}= {Q^2-M_\bk^2+|Q^2-M_\bk^2|\over 
-2Q^2(\omega+i\delta-\epsilon_\bk)}.
\end{equation}
The corresponding spectral function can be written, (at low energies) as
\begin{equation}\label{Aeff2}
A_{\rm eff}^{(1)}(\bk,\omega)= \left\{
\begin{matrix}
{Q^2-M_\bk^2\over \pi Q^2} \delta(\omega-\epsilon_\bk) & {\rm when} & M_\bk^2<Q^2 \\
0 & {\rm when} & M_\bk^2 > Q^2
\end{matrix} \right .
\end{equation}
In order to have a non-zero spectral weight near the $\Gamma$-point we require that $M_{\bk=0}^2<Q^2$, or 
\begin{equation}\label{cond1}
|\epsilon-4t|<2t_z.
\end{equation}
In addition, to obtain a single Dirac point, we require that the spectral weight is zero at $(0,\pi)$ and $(\pi,\pi)$, which translates to 
\begin{equation}\label{cond2}
|\epsilon|>2t_z, \ \ \ \ |\epsilon+ 4t| > 2t_z.
\end{equation}
These conditions are easiest to analyze in the special case when $t_z=t$. Then, a single Dirac point emerges when 
\begin{equation}\label{cond3}
2t<\epsilon<6t.
\end{equation}
We note that
this is exactly the condition for the model to be in the (1;000) topological phase,\cite{rosenberg2} i.e. STI with a surface Dirac fermion at $\bk=0$. More generally, it is easy to see that by tuning $\epsilon$ and $t_z$ one can access all 16 topological phases that can be present in a $\cT$-invariant band insulator.

In Fig.\ \ref{fig3} we plot the full spectral function $A_{\rm eff}^{(1)}(\bk,\omega)$ for a selection of parameters representing some of the topological phases. When deriving the spectral function it is essential to pick the correct complex branch of the square root in Eq.\ (\ref{geff1}). The retarded propagator that yields a positive definite spectral function and decays as $\sim 1/\omega$ at large frequencies has the following form,
\begin{equation}\label{gret}
\cG_{\rm eff, ret}^{(1)}(\bk,\omega)={g+i{\rm sgn}(\omega)\sqrt{-g^2+4Q^2[(\omega+i\delta)^2-\epsilon_\bk^2]}\over-2Q^2(\omega+i\delta-\epsilon_\bk)},
\end{equation}
with $g=(\omega+i\delta)^2-\epsilon_\bk^2+Q^2-M_\bk^2$.

\section{Interaction effects}

So far we have restricted ourselves to non-interacting Hamiltonians. As a further demonstration of the proposed models, we now study the simple case when short-range components of the intra-layer  and inter-layer   repulsive Coulomb interactions are included. We model these with the following Hubbard-like interaction terms,
\begin{equation}\label{H_int}
{\cal H}^{\rm int} =  U_1\sum_{x,\,y\, ,\ell}  n_{\ell,x} n_{\ell,y} + U_2\sum_{x,\,y}n_{1,x} n_{2,y},
\end{equation}
where $x,\,y$ label the position, spin (and orbital for model I) indices. The interaction parameters for different orbitals will generically be different but for the sake of simplicity we limit ourselves to the above two-parameter model. Furthermore, we shall only consider on-site interactions and will hereafter drop the position index. As already mentioned, the leading instability associated with each interaction is considered leading to exciton condensation for the inter-layer interaction, and the ferromagnetic instability for the intra-layer interaction. We use a simple mean-field decoupling to diagonalize the lattice models exactly and the resulting phase diagram is obtained in the $U_1$-$U_2$ plane. The calculation carries through essentially unchanged from the equivalent calculation for the continuous Dirac Hamiltonian \cite{seradjeh1} with which we compare our results. 

\subsection{Exciton instability}

When including the inter-layer Coulomb interaction, electrons and holes on opposite surfaces can form excitons and condense. The interaction can be decoupled with a matrix-valued order parameter $\Gamma=U_2 \langle \Psi_1\Psi_2^\dagger \rangle$ with the surface spinor in surface $\ell$, $\Psi_\ell$, having 4 and 2 components for model I and II respectively. Using the 8 or 4-component spinor for the combined surfaces $\Psi=(\Psi_1,\, \Psi_2)^T$, we write the expectation value taken with respect to the mean-field Hamiltonian as 
\begin{equation}\label{H_mf}
{\cal{H}}^{\textrm{EC}}_{I/II} = {\cal{H}}_{I/II} + \Psi^\dagger \begin{pmatrix} 0 & \Gamma \\ \Gamma^\dagger & 0 \end{pmatrix} \Psi + \frac{1}{U_2}\textrm{tr}(\Gamma^\dagger\Gamma).
\end{equation}  
For chemical potential $\mu$ close to zero, an order parameter that opens a gap in the spectrum lowers the overall kinetic energy and will be favored. The maximal gap is obtained by considering an order parameter that anticommutes with the kinetic part of the Hamiltonian. This form is diagonal in spin space and also preserves time-reversal symmetry. For model I we find
\begin{equation}
\Gamma_I=\begin{pmatrix}
0 & 2i\Delta\mathbb{1} \\
-2i\Delta\mathbb{1} & 0 
\end{pmatrix},\end{equation}
and for model II
\begin{equation}
\Gamma_{II}=i\Delta\mathbb{1},
\end{equation}
with $\Delta$ assumed real and $\mathbb{1}$ being the identity matrix in spin space. The mean-field Hamiltonians can be diagonalized exactly to find the following spectra
\begin{eqnarray}\label{ek_gamma_I}
E_{\bk,\, s,\, a,\, b}^{(I)}&=& -\mu + a\bigg[\epsilon_\bk^2+V^2+\bar{M}_\bk^2+2\Omega_\bk^2 + 4\Delta^2  \nonumber \\ 
&+&2b \sqrt{(s \epsilon_\bk V - \Omega_\bk^2 )^2 + \bar{M}_\bk^2(V^2+\Omega_\bk^2)}\bigg]^{1/2}\!\!\!,\,\:\:\:\:\:
\end{eqnarray}
and
\begin{equation}\label{ek_gamma_II}
E_{\bk,\,s,\,a}^{(II)}= -\mu + a \sqrt{(V + s \epsilon_\bk)^2 + \bar{M}_\bk^2 +\Delta^2},
\end{equation}
with $s,a,b=\pm1$. % and where $s$ and $b$ label the spin and orbital degree of freedom while $a$ labels the positive and negatives energy branches. 
The relation between $\Delta$ and $U_2$ can then be obtained from the gap equations which result from minimizing the ground state energy
\begin{equation}
E_g^{(I)} = {\sum}_{\bk,\, s,\, a,\, b}' E_{\bk,\, s,\, a,\, b}^{(I)} + 16\frac{\Delta^2}{U_2}N,
\end{equation}
and 
\begin{equation}
E_g^{(II)} ={\sum}_{\bk,\, s,\, a}' E_{\bk,\, s,\, a}^{(II)} + 2\frac{\Delta^2}{U_2}N,
\end{equation}
where $\sum'$ indicates a sum over occupied states and $N$ is the number of sites of the 2D lattice. The resulting gap equations are
\begin{equation}\label{gap_I_TEC}
\Delta = -\frac{U_2}{8N}{\sum}_{\bk,\, s,\, a,\, b}' \frac{\Delta}{E_{\bk,\, s,\, a,\,b}^{(I)} +\mu},
\end{equation}
and 
\begin{equation}\label{gap_II_TEC}
\Delta = -\frac{U_2}{4N}{\sum}_{\bk,\, s,\, a}' \frac{\Delta}{E_{\bk,\, s,\, a}^{(II)} +\mu}.
\end{equation}

\begin{figure*}[hbt]
\includegraphics[width = 9.0cm]{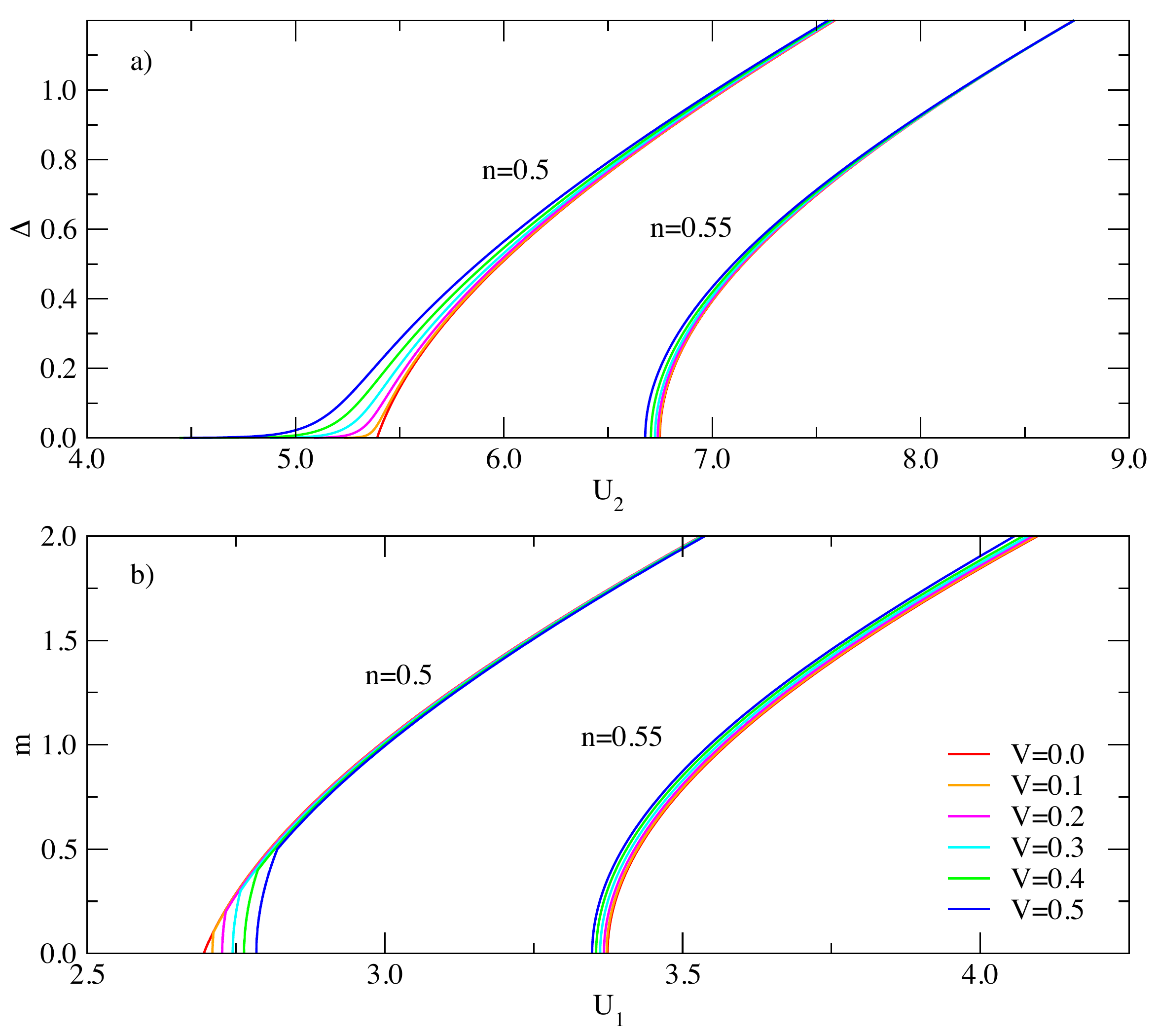}\includegraphics[width = 9.0cm]{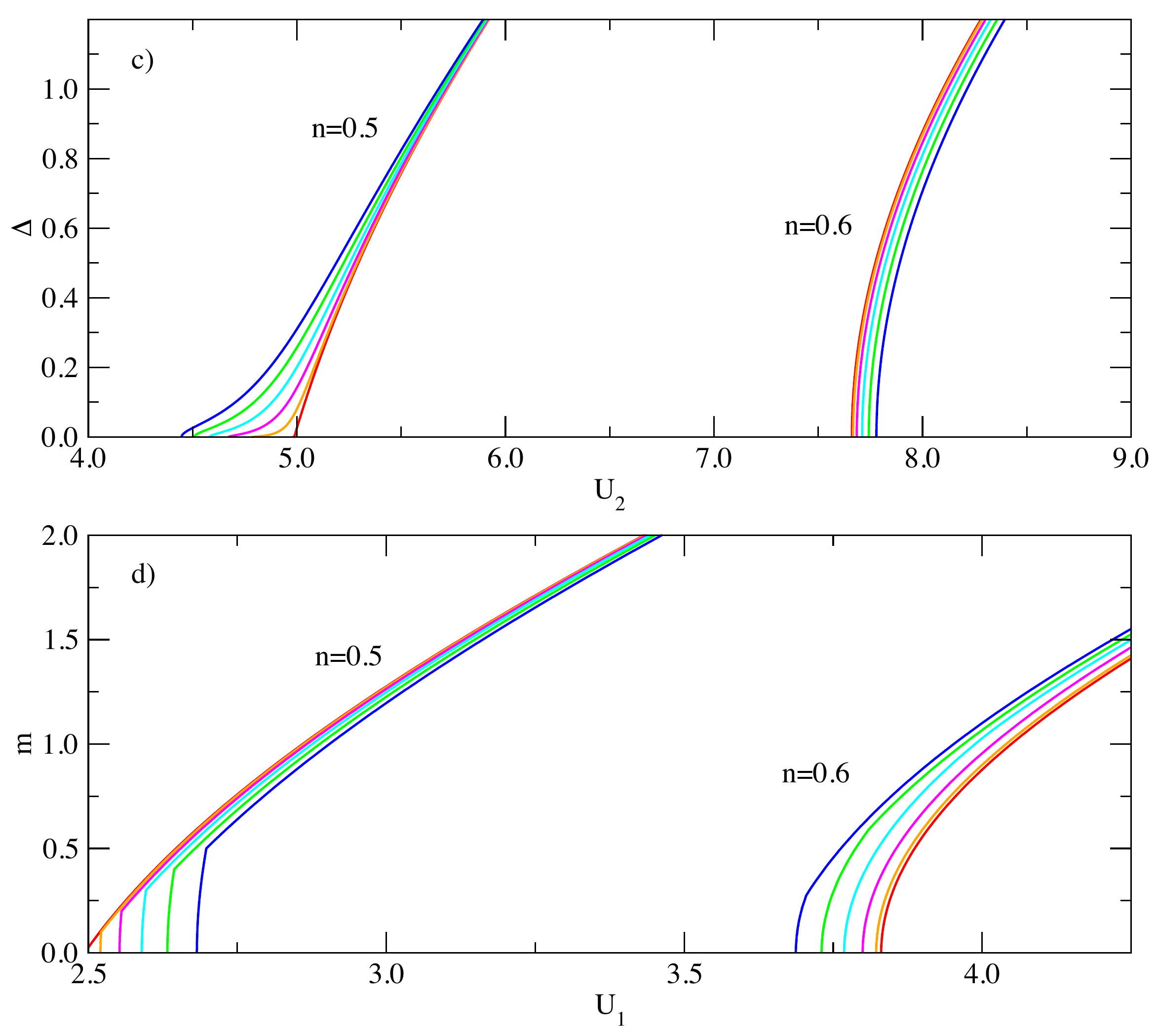}
\caption{Gap associated with the EC and magnetic instability as a function of the strength of the short-range Coulomb interaction for Hamiltonian I (left column) for filling fractions $n=0.5,\,0.55$  and Hamiltonian II (right column) for filling fractions $n=0.5,\,0.6$. Panels (a) and (c) show half the EC gap as a function of the inter-layer interaction $U_2$ and panels (b) and (d) show half the magnetic gap as a function of intra-layer interaction $U_1$. Results obtained by solving the mean-field gap equations (\ref{gap_I_TEC}), (\ref{gap_II_TEC}), (\ref{gap_I_FM}) and (\ref{gap_II_FM}) for a finite system of 1000 by 1000 sites. 
}\label{fig4}
\end{figure*}

These gap equations can be solved numerically for a finite, but large enough system (here a square lattice of 1000 by 1000 sites) with a fixed number of electrons, or fixed filling fraction, as opposed to a fixed chemical potential. The number of electrons in a calculation is given by $8L^2n$ and $4L^2n$ for model I and II respectively, with $n\in[0,\,1]$ being the filling fraction and $L^2$ the number of sites. We present results for the half-filled case $n=0.5$ and for a slightly higher filling with $n=0.55$ for model I and $n=0.6$ for model II, such that we are comparing both models for the same number of electrons in the Dirac bands. Figures \ref{fig4} (a,c)  show half the size of the gap (the actual gap being $2\Delta$) associated with the EC instability of each model.

Figures \ref{fig4} (a,c) help illustrate an important point regarding the exciton condensate in this system. For $V\neq 0$ and zero chemical potential the latter formally occurs at an infinitesimal coupling $U_2$ because the relevant gap equations are of the BCS form, familiar from the theory of superconductivity. (At $V=0$ the density of states at the Fermi level $N_F$ vanishes so the formation of EC requires a finite critical coupling $U_{2c}$.) Nevertheless we observe from Figs.\ \ref{fig4} (a,c) that for $U_2<U_{2c}$ the exciton gap rapidly vanishes even for a relatively large bias $V=0.5$, consistent with the BCS exponential form $\Delta\simeq e^{-1/U_2N_F}$. Thus, as a practical matter, it would appear that biasing the system does not significantly improve the chances for experimentally observing the exciton condensate, unless the interaction strength can be tuned to the vicinity of $U_{2c}$. We also emphasize that the precise numerical value of $U_{2c}$ obtained in this work should not be taken as a reliable guide for experiments because of our reliance on the simple MF theory and our neglect of the long-range part of the Coulomb repulsion. Accurate determination of the critical coupling and the relevant critical temperature $T_c$ is a problem of considerable complexity with predictions for the latter ranging from mK to room temperature.\cite{lozovik1,min1,efetov1,pesin1} Although the MF calculation employed in this study cannot reliably predict the absolute magnitude of the exciton gap or $T_c$, we expect that it is capable of indicating trends, as discussed above, and help in understanding the competition with other ordered phases, such as the surface magnetism discussed below.

\subsection{Ferromagnetic instability}

We mentioned that a magnetic moment on one surface  along the $z$ direction will open a gap in the Dirac cone associated with this surface by breaking time-reversal symmetry. Such a gap can also appear spontaneously due to intra-layer Coulomb interaction. We again seek a matrix order parameter that will decouple this interaction. We consider here the specific case where the Coulomb interaction has the same strength on both surfaces such that both surfaces are in the same phase and the magnetic gap will open simultaneously in all Dirac cones. The order parameter will now involve only terms diagonal in the surface inde.
% $\Upsilon = U_1 [\langle \Psi_1 \Psi_1^\dagger\rangle + \langle \Psi_2 \Psi_2^\dagger\rangle]$. 

\begin{figure*}[hbt]
\includegraphics[width = 9.0cm]{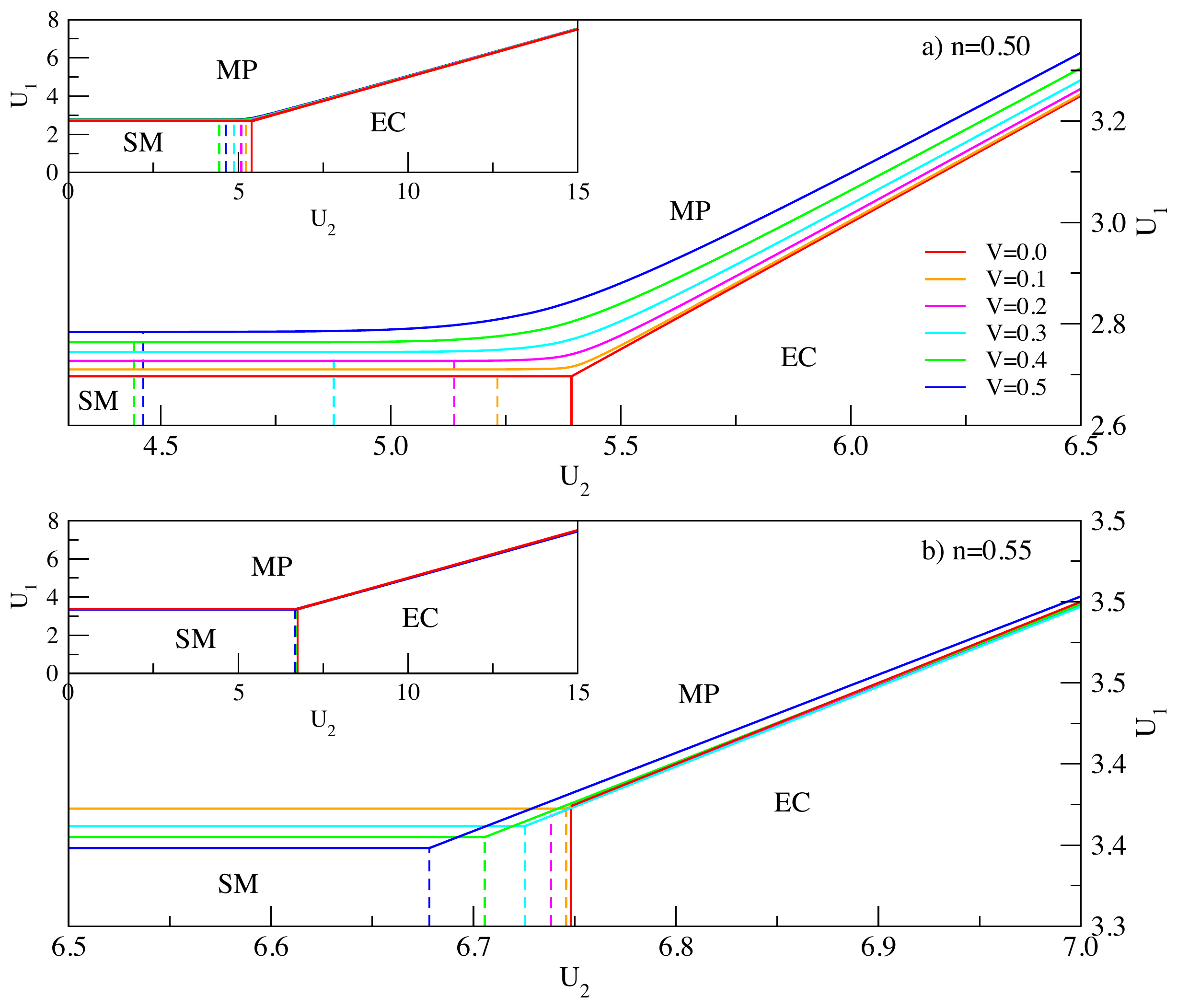}\includegraphics[width = 9.0cm]{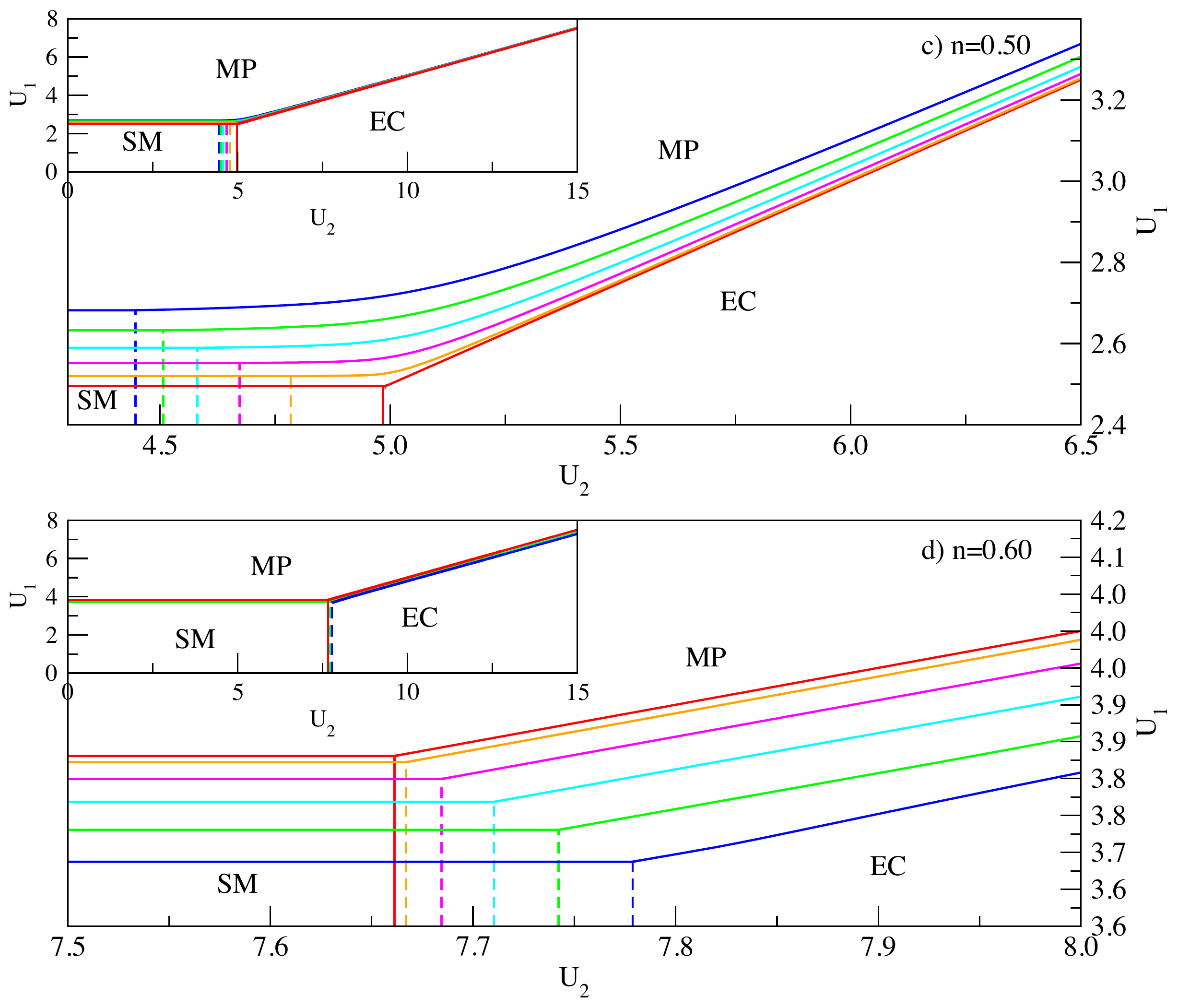}
\caption{Phase diagram as a function of the intra-layer ($U_1$) and inter-layer ($U_2$) short-range Coulomb interaction strength for Hamiltonian I (left column) and Hamiltonian II (right column). Main panels show true main region of interest while the insets sketch the full phase diagram. Panel (a) and (c) show the phase diagram for half-filling. Panel (b) shows model I at $n=0.55$ and panel (d) shows the phase diagram of model II at $n=0.6$. The filling fraction of (b) and (d) are chosen to have the same number of electron in the Dirac bands. Results obtained by solving the gap equations (\ref{gap_I_TEC}), (\ref{gap_II_TEC}), (\ref{gap_I_FM}) and (\ref{gap_II_FM}) for a finite system of 1000 by 1000 sites. 
}\label{fig5}
\end{figure*}
We start with the more complicated case of Hamiltonian I. For simplicity we will consider the Coulomb interaction to be of the same strength for all combinations of quantum numbers (layer, orbital and spin). Using $n$ as the number operator,
\begin{eqnarray}\label{H_int_I}
{\cal{H}}_I^{\rm int} = U_1 \sum_\ell \bigg[ n_{\ell,\,1,\,\uparrow} n_{\ell,\,1,\,\downarrow} &+& n_{\ell,\,2,\,\uparrow} n_{\ell,\,2,\,\downarrow} \nonumber\\
&+&  \sum_{s,\,t} n_{\ell,\,1,\,s} n_{\ell,\,2,\,t} \bigg],
\end{eqnarray}  
with the spin degrees of freedom $s,\,t\in \{ \uparrow,\, \downarrow\}$, $1$ and $2$ labeling the orbitals and $\ell$ being the surface index as before. We rewrite the mean-field version of this Hamiltonian by defining the average magnetization within an orbital $o$ as the imbalance between spin up and down electrons
\begin{equation}\label{m}
m_{\ell,\,o}=\frac{U_1}{2}\big[\langle n_{\ell,\,o,\,\uparrow}\rangle-\langle n_{\ell,\,o,\,\downarrow}\rangle\big],
\end{equation}  
and the average number of electrons in an orbital
\begin{equation}\label{n}
\bar{n}_{\ell,\,o}=\frac{U_1}{2}\big[\langle n_{\ell,\,o,\,\uparrow}\rangle+\langle n_{\ell,\,o,\,\downarrow}\rangle\big].
\end{equation}  
The resulting mean-field Hamiltonian is
\begin{eqnarray}\label{H_mf_mag}
{\cal{H}}^{\textrm{FM}}_{I} &=& {\cal{H}}_{I} + \frac{1}{U_1}\sum_\ell \Bigg[ \big( m_{\ell,\,1}^2- \bar{n}_{\ell,\,1}^2 + m_{\ell,\,2}^2 - {\bar{n}}_{\ell,\,2}^2\big) \nonumber\\
&+&\Psi_\ell^\dagger\begin{pmatrix} C_\ell\mathbb{1} - m_{\ell,\,1} s_z & 0 \\ 0 & D_\ell\mathbb{1} - m_{\ell,\,2} s_z \end{pmatrix}\Psi_\ell\Bigg],
\end{eqnarray}
with
\begin{equation}
C_\ell = [\bar{n}_\ell+\bar{n}_{\ell,\,2}] \quad  \textrm{and} \quad D_\ell =  [\bar{n}_\ell+\bar{n}_{\ell,\,1}],
\end{equation}
where we defined $\bar{n}_\ell=\bar{n}_{\ell,\,1}+\bar{n}_{\ell,\,2}$. The part of the Hamiltonian proportional to $\bar{n}_\ell$ is just a shift in the chemical potential at each surface and can be dropped here. At exactly half-filling and zero bias, $\bar{n}_{\ell,\,1}=\bar{n}_{\ell,\,2}$ such that $C_\ell$, $D_\ell$ and terms proportional to $\bar{n}_{\ell,\,1}^2$ and $\bar{n}_{\ell,\,2}^2$ can all be similarly ignored. Since we assumed the same interaction in both bands, we get a maximum gap opening when $|m_{\ell,\,1}|=|m_{\ell,\,2}|=m$ and the order parameter is simply
\begin{equation}\label{Upsilon_I}
\Upsilon_I=m\begin{pmatrix}
s_z & 0 \\
0 & -s_z 
\end{pmatrix}.\end{equation}
This form can also be obtained by requiring that the order parameter anticommutes with $H_I$ for $V=0$. This holds for any filling fraction as the anti-commutation relation does not depend on this parameter, but $m$ loses its meaning as the average magnetization for each orbital away from half-filling. For non-zero biases, no order parameter that anticommutes with the Hamiltonian can be found and it is therefore possible that a different form will be preferred. We shall nonetheless use this same form even for finite bias for the sake of simplicity. As noted before, Eq. (\ref{H_mf_mag}) can now be rewritten by replacing $h_\bk\to h_\bk+m s_z$ with the dispersion relation remaining the same as Eq. (\ref{ek_gamma_I}), but with $\epsilon_\bk^2=4(\sin^2{k_x}+\sin^2{k_y})+m^2$. The ground state energy and the resulting gap equation are
\begin{equation}
E_g^{(I)} = {\sum}_{\bk,\, s,\, a,\, b}' E_{\bk,\, s,\, a,\, b}^{(I)} + 2\frac{m^2}{U_1}N,
\end{equation}
and
\begin{eqnarray}\label{gap_I_FM}
m = -\frac{U_1}{4N}&&{\sum}_{\bk,\, s,\, a,\, b}'  \frac{m}{E_{\bk,\, s,\, a,\,b}^{(I)} +\mu}\\\nonumber
&\times&\Bigg[1+\frac{bV(V-s\Omega_\bk^2/\epsilon_\bk)}{\sqrt{(s \epsilon_\bk V - \Omega_\bk^2 )^2 + \bar{M}_\bk^2(V^2+\Omega_\bk^2)}}\Bigg].\,\,
\end{eqnarray}

Model II with its single orbital is much simpler. Following the same procedure as for model I we find that the interaction term
\begin{equation}\label{H_int_II}
{\cal{H}}_{II}^{\rm int} = U_1 \sum_\ell  n_{\ell,\,\uparrow} n_{\ell,\,\downarrow},
\end{equation}  
can be decoupled for zero bias with the order parameter 
\begin{equation}\label{Upsilon_II}
\Upsilon_{II}=m s_z, 
\end{equation}
which anticommutes with $H_{II}$. $m$ now preserves its meaning as the average magnetization for any filling fraction at zero bias. We chose again to use the same order parameter even at finite bias for simplicity even if a different form could be favored. Using the same substitution for $\epsilon_\bk$, the spectrum is still given by Eq. (\ref{ek_gamma_II}) while the ground state energy is
\begin{equation}
E_g^{(II)} = {\sum}_{\bk,\, s,\, a}' E_{\bk,\, s,\, a}^{(II)} + \frac{m^2}{U_1}N,
\end{equation}
and the gap equation
\begin{equation}\label{gap_II_FM}
m = -\frac{U_1}{2N}{\sum}_{\bk,\, s,\, a,\, b}'  \frac{m}{E_{\bk,\, s,\, a}^{(II)} +\mu}(1+sV/\epsilon_\bk).
\end{equation}

The magnetic gap equations obtained are solved numerically in the same fashion as for the EC instability and panels (b) and (d) of Fig. \ref{fig4} show half the size of the gap (the actual gap being $2m$).

\subsection{Phase diagram}

\begin{figure*}[hbt]
\includegraphics[width = 9.0cm]{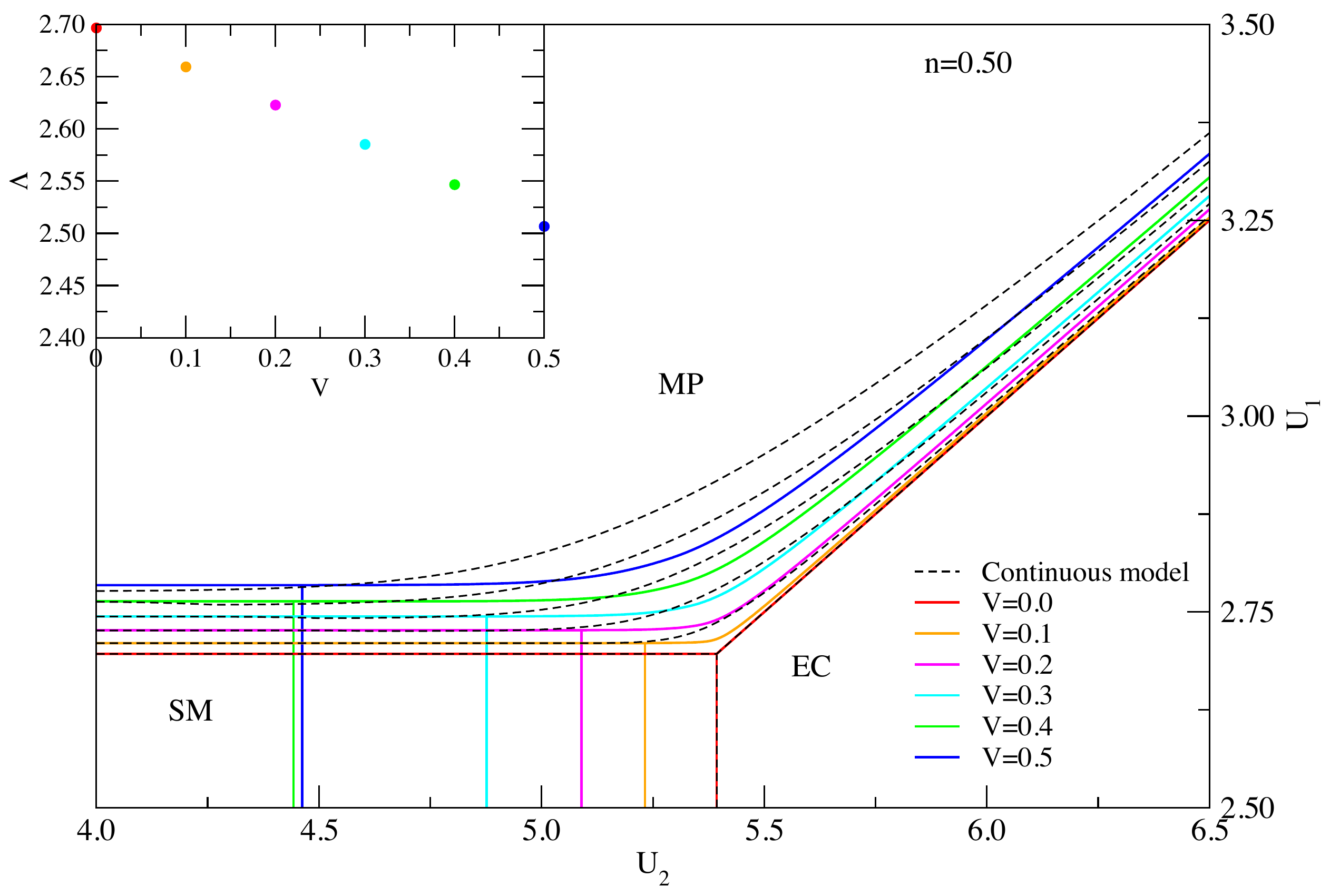}\includegraphics[width = 9.0cm]{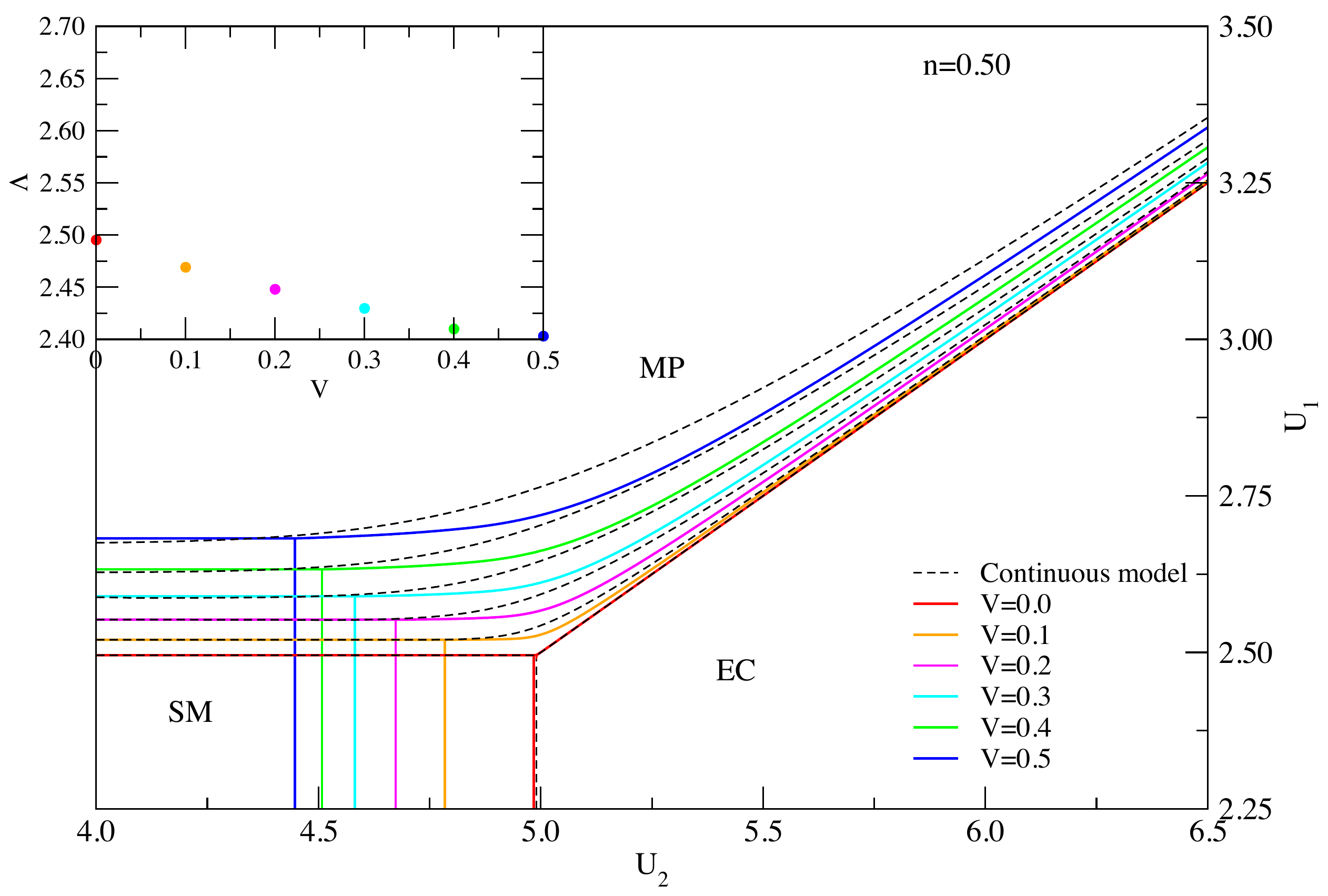}
\caption{Comparison of the phase diagram for the continuous Dirac Hamiltonian (\ref{dir}) and the lattice Hamiltonian I in panel (a) and Hamiltonian II in panel (b). This phase diagram is for the half-filling case. The momentum cutoff $\Lambda$ used for the continuous model (shown in the insets) is tuned to reproduce the critical intra-layer coupling $U_{1c}$ obtained from the lattice models I and II. Results obtained by solving the gap equations (\ref{gap_I_TEC}), (\ref{gap_II_TEC}), (\ref{gap_I_FM}) and (\ref{gap_II_FM}) for a finite system of 1000 by 1000 sites. 
}\label{fig6}
\end{figure*}

We can now draw the phase diagram of a STI as a function of the strength of both the intra-layer and inter-layer interaction. The phase boundary between the magnetic phase (MP) and the EC is calculated by evaluating the ground state energy and the coupling strength as a function of the gap simultaneously to obtain the functions $U_1(E_g|_{\Delta=0})$ and $U_2(E_g|_{m=0})$ and plotting the resulting parametric curve in the $U_1$-$U_2$ plane. Fig. \ref{fig5} shows the resulting phase diagram in detail while Fig. \ref{fig6} compares model I and II to the continuous Dirac Hamiltonian (\ref{dir}). 

It should be noted that all models show a critical intra-layer coupling below which we are in a non-magnetic phase. For the EC however such a critical inter-layer coupling should only exist for zero bias and otherwise vanish. Fig. \ref{fig4} and \ref{fig5} show that the lattice models instead always have a critical inter-layer coupling (shown as dashed lines) even at non-zero bias. Below both critical couplings we are in an uncondensed, or semi-metallic (SM) phase. This is an artifact of the lattice model due to a very small gap opening at $\mu=0$ for non-zero bias. The size of this gap increases slower than $\sim 0.1V^3$ for model I and $\sim 0.25V^2$ for model II in the parameter range considered here. This only becomes relevant when studying the weakly interacting case where the continuous model predicts an exponentially small gap, and therefore difficult to observe experimentally. In the region of interest for experiments and possible devices, this small gap can be ignored safely compared to the EC gap.

Fig. \ref{fig6} with its comparison of model I and II to the continuous Dirac Hamiltonian (\ref{dir}) raises a number of interesting points. Although the low-energy physics of the surface states of a STI is generally thought to be well described by this effective continuous model, a comparison to our own effective lattice model suggests that it will be insufficient for a quantitative study of real TI materials. Not only does (\ref{dir}) depend on an arbitrary cutoff, the exact position of the phase boundary differs significantly at intermediate couplings. This means that estimates derived from the Dirac Hamiltonian for the critical temperature of the EC, for example, are unlikely to be very accurate. A more ambitious study with any hope of achieving a quantitative estimate will require a model such as those proposed herein, that accounts properly for the underlying lattice structure of the STI and its more complicated band structure.

Despite its limited applicability, the mean-field calculation above also allows us some general observations of use for the experimental observation of EC. Fig. \ref{fig5} suggests for example that an appropriate system for the observation of a EC requires an inter-layer coupling at least twice as strong as the intra-layer coupling, otherwise the magnetic phase takes over as the most favorable state and the topological properties of the surface states are lost due to the breaking of time-reversal symmetry.

\subsection{Comparison to 3D results}
To ascertain the accuracy of the proposed models in terms of predicting the correct phase diagrams we now compare the results obtained above to a calculation based on the exact surface propagator derived in Sec.\ III. We remark that employing the exact surface propagator is equivalent to performing the calculation within the full 3D lattice model. 

To this end it is useful to reformulate the mean field theory in terms of the path integral of Sec.\ III.
Thus, we write the action for the 3D interacting TI as
\begin{equation}\label{sint}
S^{\rm int}= S_0+\int_0^\beta d\tau {\cal H}^{\rm int}
\end{equation}
where $S_0$ is the noninteracting action defined by Eq.\ (\ref{s1}) and  ${\cal H}^{\rm int}$ represents the Hubbard term (\ref{H_int}). For simplicity we consider only the on-site interaction $U_1$ and neglect interactions between different orbitals. On every site the interaction term can be decoupled using the Hubbard-Stratonovich transformation based on the identity
\begin{equation}\label{strat}
e^{{1\over 2}U_1(n_\uparrow-n_\downarrow)^2}=C\int{\cal D}m \,e^{-{1\over 2U_1}m^2-m(n_\uparrow-n_\downarrow)}
\end{equation}
where $C$ is a constant and $m$ represents an auxiliary real field. Noting further that $(n_\uparrow-n_\downarrow)^2=n_\uparrow+n_\downarrow-2n_\uparrow n_\downarrow$ one can express the Hubbard term in terms of the total charge and the total magnetization on each site, thus rendering the action (\ref{sint}) formally bilinear in Fermionic fields,
\begin{equation}\label{sint2}
S[m]= S_0+\int_0^\beta d\tau \sum_\alpha\left[{m_\alpha^2(\tau)\over 2U_1}+m_\alpha(\tau)(n_{\alpha\uparrow}-n_{\alpha\downarrow})\right],
\end{equation}
where $\alpha$ represents the spatial coordinates and orbital indices. The  $(n_\uparrow+n_\downarrow)$ term from the decoupled ${\cal H}^{\rm int}$ has been absorbed into the chemical potential term.

The mean field approximation consists of neglecting the temporal (quantum) fluctuations in $m_\alpha(\tau)$, which can then be interpreted as on-site magnetization. If we furthermore assume a homogeneous solution, i.e. $m_\alpha=m$ for all sites and orbitals, we can pass to the Fourier space and the magnetization term is then seen to modify the non-interacting action $S_0$ by replacing $h_\bk\to h_\bk +m s_z$. The equilibrium magnetization then follows from minimizing the effective action $S_{\rm eff}[m]$
with respect to $m$. The former is obtained by integrating out the Fermi fields in $S[m]$, 
\begin{equation}\label{sm}
e^{-S_{\rm eff}[m]}=\int{\cal D}[\bar{\chi},\chi;\bar{\Psi},\Psi]e^{-S[m]}.
\end{equation}
We proceed in two steps. First we integrate out the bulk degrees of freedom and focus on the surfaces. The procedure is identical to the one outlined in Sec.\ III. For a semi-infinite slab geometry it leads to the effective surface action given in terms of the surface propagator ${\cal G}_{\rm eff}^{(1)}$ defined in Eq.\ (\ref{geff1}) with $\epsilon_\bk=s\sqrt{4(\sin^2{k_x}+\sin^2{k_y})+m^2}$, where $s=\pm$ now labels the two spin projections expressed in the helical basis. The effective MF action is thus given by 
\begin{eqnarray}\label{smf}
&& e^{-S_{\rm eff}[m]}= e^{-\beta{m^2\over U_1}N} \int{\cal D}[\bar{\Psi},\Psi]\times\\
&& 
\exp{\left\{-{1\over\beta} \sum_{n,\bk,s}\bar{\Psi}_s(i\omega_n,\bk)\frac{1}{\cG_{\rm eff}^{(1)}(i\omega_n,\bk,s)}\Psi_s(i\omega_n,\bk)\right\} }. \nonumber
\end{eqnarray}
The integral over the remaining Fermi fields is easily evaluated and leads to 
\begin{eqnarray}\label{smf2}
S_{\rm eff}[m]= \beta{m^2\over U_1}N+
\sum_{n,\bk,s} \ln\left[\cG_{\rm eff}^{(1)}(i\omega_n,\bk,s)\right].
\end{eqnarray}

Before tackling the minimization it useful to first perform the summation over the spin index which yields a more transparent expression
\begin{eqnarray}\label{smf3}
S_{\rm eff}[m]= \beta{m^2\over U_1}N+
\sum_{n,\bk} \ln\left({D_n^2\over \omega_n^2+\epsilon_\bk^2}\right),
\end{eqnarray}
with $D_n= \cG_{\rm eff}^{(1)}(i\omega_n,\bk)H_+$.
\begin{figure}[t]
\includegraphics[width = 8cm]{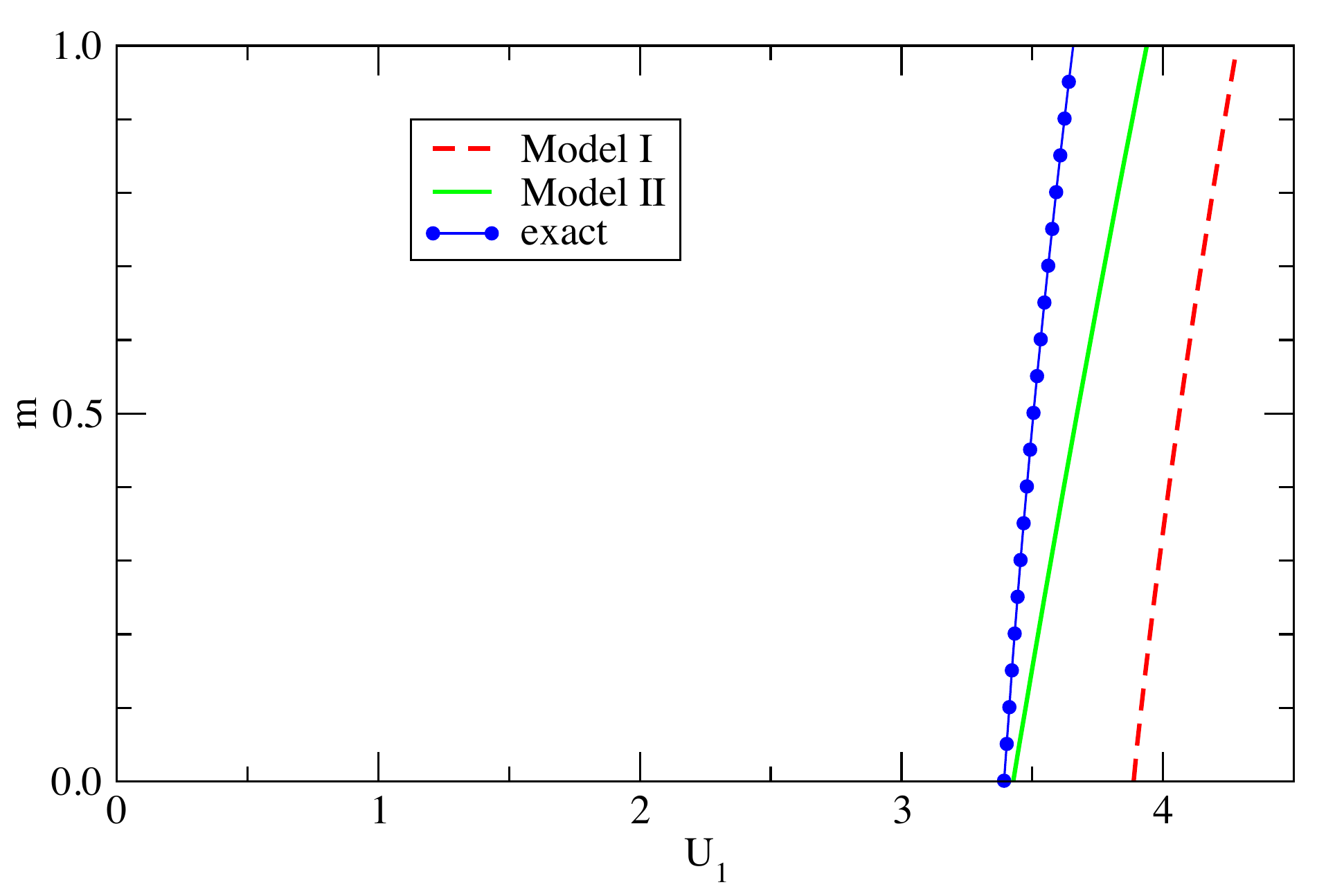}
\caption{Comparison of model I, II with the exact result for the dependence of magnetization $m$ on the interaction parameter $U_1$. The parameters are $t_z=t=\lambda_z=1.0$, $\epsilon=4.0$ which corresponds to the bulk STI in (1;000) phase and $V=\mu=0$.
}\label{fig7}
\end{figure}
Differentiating the result in Eq.\ (\ref{smf3}) with respect to $m$ leads to the exact gap equation (which we omit here for the sake of brevity). It is a simple matter to 
solve this gap equation numerically and compare the outcome to our previous results based on the surface-only models I and II. The dependence of $m$ on $U_1$ is shown in Fig.\ \ref{fig7}. We observe that the agreement is quite good, especially for model II whose critical coupling $U_{1c}^{II}$ falls within 2\% of the exact value $U_c=3.39$ for the model parameters specified in the caption of Fig.\ \ref{fig7}. We have considered other values of model parameters and found similar level of agreement, with the model I and II critical couplings lying typically within 10-15\% of the exact value. In all cases we found that model I and II critical couplings exceed the exact value.

%%%%%%%%%%%%%%%%%%%%%%%%%%%%%%%%%%%%%%%%%%%%%%%%%%%%%%%%%%%%%%%%%%%%%%%

\section{Concluding remarks}

The key advance presented in this work consists of the formulation of two simple 2D lattice models that faithfully describe the low-energy electronic states associated with a pair of surfaces of a strong topological insulator.
We have demonstrated that these models produce electron states with properties characteristic of true STI surface states, in terms of their spectral properties as well as their susceptibility to the formation of ordered phases in the presence of interactions. Employing these surface-only models will confer a significant advantage on any numerical calculation by obviating the need to explicitly consider the (uninteresting) bulk degrees of freedom. This includes computations that seek to address the effects of interactions (both electron-electron and electron-phonon) as well as those of various types of disorder or other perturbations, such as magnetism or superconductivity, acting on the STI surface. 

Although we have focused here on the example of two parallel surfaces one can imagine applying the same strategy to other geometries and situations. One example that we have considered (but not discussed here in detail) is the surface of a cylindrical STI wire which can be modeled by 2D lattice Hamiltonians similar to those defining our model I or II with an extra hopping allowed between radially opposed sites on the cylinder. Also, one can straightforwardly apply this strategy to 2D topological insulators (as well as 2D Chern insulators) and formulate 1D lattice models for the pair of edges. Finally, one can imagine constructing lattice models for 2D surfaces or 1D edges for 3D or 2D topological superconductors using our techniques since their first-quantized, single-particle Hamiltonians are in many cases identical to those describing TIs.  

\section{Acknowledgments}
The authors are indebted to Y.P. Chen,  A. Cook, G. Rosenberg, H. Omid and C. Weeks for insightful discussions. The work reported here has been supported by DARPA through the MESO-TI initiative (grant N66001-11-1-4107).

\end{document}